\documentclass[11pt, fleqn]{amsart}
\usepackage{bm}
\usepackage{multirow}
\usepackage{graphicx}
\usepackage{amsmath}
\usepackage{mathtools}
\usepackage[foot]{amsaddr}
\usepackage{natbib}
\usepackage{float}
\usepackage{color}
\usepackage[table]{xcolor}
\usepackage{algorithm,algorithmicx}
\usepackage{algpseudocode}

\makeatletter
\renewcommand*{\env@matrix}[1][*\c@MaxMatrixCols c]{%
  \hskip -\arraycolsep
  \let\@ifnextchar\new@ifnextchar
  \array{#1}}
\makeatother

\usepackage{todonotes}

\theoremstyle{definition}

\theoremstyle{remark}

\numberwithin{equation}{section}

\DeclareMathOperator{\cor}{corr\,}

\DeclareMathOperator{\caa}{\widehat{\cor}(\textit{Y}_1,\widetilde{\textit{Z}}_1)}
\DeclareMathOperator{\cab}{\widehat{\cor}(\textit{Y}_1,\widetilde{\textit{Z}}_2)}
\DeclareMathOperator{\cba}{\widehat{\cor}(\textit{Y}_2,\widetilde{\textit{Z}}_1)}
\DeclareMathOperator{\cbb}{\widehat{\cor}(\textit{Y}_2,\widetilde{\textit{Z}}_2)}

\DeclareMathOperator{\xba}{\widehat{\cor}(\textit{X}_7,\widetilde{\textit{Z}}_1)}
\DeclareMathOperator{\xbb}{\widehat{\cor}(\textit{X}_7,\widetilde{\textit{Z}}_2)}

\bibpunct[, ]{(}{)}{;}{a}{}{,} 

\begin{document}

\title[]{Data Fusion for Joining Income and Consumption Information Using Different Donor-Recipient Distance Metrics}
\author{Florian Meinfelder}
\address{Florian Meinfelder \\ Lehrstuhl f\"{u}r Statistik und \"Okonometrie \\ Feldkirchenstra{\ss}e 21 \\ 96052 Bamberg \\
Germany} \email{florian.meinfelder@uni-bamberg.de}

\author{Jannik Schaller}
\address{Jannik Schaller \\ Statistisches Bundesamt (Destatis) \\ Gustav-Stresemann-Ring 11 \\ 65189 Wiesbaden \\ Germany} \email{jannik.schaller@destatis.de}

\thanks{\renewcommand{\baselinestretch}{1}\scriptsize The authors would like to thank Eurostat for the permission to use their \texttt{R} code. We also thank Pierre Lamarche to share his unpublished article, which is currently under review. The corresponding author would like to thank the Federal Statistical Office of Germany (Destatis) as well as the University of Trier for financial support.}

\renewcommand{\baselinestretch}{2}\normalsize

\begin{abstract}
Data fusion describes the method of combining data from (at least) two initially independent data sources to allow for joint analysis of variables which are not jointly observed. The fundamental idea is to base inference on identifying assumptions, and on common variables which provide information that is jointly observed in all the data sources. A popular class of methods dealing with this particular missing-data problem is based on nearest neighbour matching. However, exact matches become unlikely with increasing common information, and the specification of the distance function can influence the results of the data fusion. In this paper we compare two different approaches of nearest neighbour hot deck matching: One, Random Hot Deck, is a variant of the covariate-based matching methods which was proposed by Eurostat, and can be considered as a 'classical' statistical matching method, whereas the alternative approach is based on Predictive Mean Matching. We discuss results from a simulation study to investigate benefits and potential drawbacks of both variants, and our findings suggest that Predictive Mean Matching tends to outperform Random Hot Deck.
\end{abstract}

\keywords{\renewcommand{\baselinestretch}{1}\scriptsize Statistical Matching, Missing Data, Predictive Mean Matching, Nearest Neighbour Imputation, Missing-By-Design Pattern}%

\maketitle

\newpage

\section{Introduction}

Data fusion, also known as statistical matching, is a perfect example of secondary data analysis. The objective of a data fusion is to jointly analyse variables from (at least) two different data sources which were not jointly observed, and each of the data sources originally served a different purpose. 

The studies of the National Statistical Institutes (NSIs)  are often committed to a particular objective such as measuring, e.g. consumption expenditure of private households, in great detail. If the need arises to incorporate and combine information from several objectives, data fusion is a standard mean to provide a microdata source, where these different types of information are artificially joined on an individual or household level. 

In 2009, the Stiglitz-Sen-Fitoussi commission \citep{Stiglitz.2009} published a report on welfare and its components, which led to various approaches among NSIs within the European Union to measure the dimensions 'income', 'consumption' and 'wealth' (ICW) as proposed by the commission. Countries which do not measure all three dimensions within a single official statistics data source have been exploring data fusion methods in order to provide a corresponding data base \citep[see e.g.][]{Donatiello.2014,Ucar.2016,Albayrak.2017,Dalla.2019}. The proposed data fusion methods are largely based on the research conducted by the European Statistical Office (Eurostat) \citep{Lamarche.2018}\footnote{This paper is currently under review and, therefore, unpublished yet. A preliminary and freely accessible version of this paper is \citet{Lamarche.2017}.} and several NSIs \citep[see e.g.][]{DOrazio.2018} on statistically matching data from EU-SILC with data from the Household Budget Survey (HBS). By matching EU-SILC and HBS, Eurostat and the NSIs pursue the goal to provide joint information about the household income details observed from EU-SILC and the household consumption expenditures observed from HBS. The original focus of their analyses had been on preserving marginal distributions and one of the preliminary findings is that Random Hot Deck (RHD), a classical nearest neighbour matching technique, performs very good in terms of preserving marginal distributions \citep{Webber.2013,Tonkin.2017,Lamarche.2018}. However, the preservation of marginal distributions does not give any hint about whether the joint distributions of the variables not jointly observed, income and consumption expenditures in our case, is adequately reproduced in the matched data file.

Our research connects by extending the analysis objective to investigating associations between different variables, and we will explain in the following, why RHD yields good results for marginal distributions, but not necessarily for conditional or joint distributions. We investigate the properties of an imputation method called Predictive Mean Matching (PMM) \citep{Rubin.1986} which was extended by \citet{Little.1988} to multivariate data situations (which is a typical scenario for data fusion settings). While both, RHD and PMM, are based on nearest neighbour hot deck matching, the underlying principles are very different, as RHD matches on the combined distances of the covariates, jointly observed in both studies, whereas PMM matches on the combined distances of model-based predictions of the variables which are observed in only one of the two studies. While these two methods do not exhaust the plethora of available nearest neighbour matching variants, they can be considered as archetypical for what we like to refer to as covariate-based matching and model-based matching. In order to discuss benefits and drawbacks of both data fusion archetypes, we compare the performance of Random Hot Deck and Predictive Mean Matching within a simulation study. As an extension to the research conducted by Eurostat and the NSIs, we focus on preserving joint distributions and associations between different variables, especially between the variables not jointly observed.

We focus in this paper on point estimators and descriptive statistics, and we therefore do not address incorporating uncertainty caused by the originally missing data, but extension to Multiple Imputation for multivariate PMM is straightforward \citep[see e.g.][]{Rubin.1986, KollerMeinfelder.2009}.

Nearest neighbour methods also spawn various other interesting research problems, such as balancing the usage of donors using constrained matching \citep[see e.g.][]{Rodgers.1984,Rubin.1986} or selecting donors from $k$ Nearest Neighbours \citep[see e.g.][]{Chen.2000,Andridge.2010,Beretta.2016}. For the sake of simplicity, we do not explicitly address these issues in the following, although the setup of our simulations could be extended accordingly.

To meet our objective of comparing the data fusion performance of RHD and PMM, we structure our paper as follows: Section \ref{sec:MethAsp} contains a general overview of data fusion, followed by an in-depth description of the two aforementioned algorithms in Section \ref{sec:Algo}. We investigate the properties of RHD and PMM as data fusion methods within a simulation study based on Scientifc Use Files (SUFs) data from the EU-SILC study which we modify to mimic a data fusion situation. The setup of this stimulation study is described in Section \ref{sec:Sim_Design} and we discuss the corresponding results in Section \ref{sec:results}. We conclude the findings of our research in the final Section \ref{sec:Conclusion}.

\vspace{20mm}

\section{Methodological Aspects of Data Fusion Scenarios}\label{sec:MethAsp}

In this section we introduce the basic notation used throughout this paper, and we introduce perspectives on data fusion from statistical literature as well as the practitioners' perspective which often relies on the identification of 'statistical twins' or nearest neighbours.

\subsection{Theoretical Background}

\begin{figure}[tbh] 
    \centering
    \includegraphics[width=7cm]{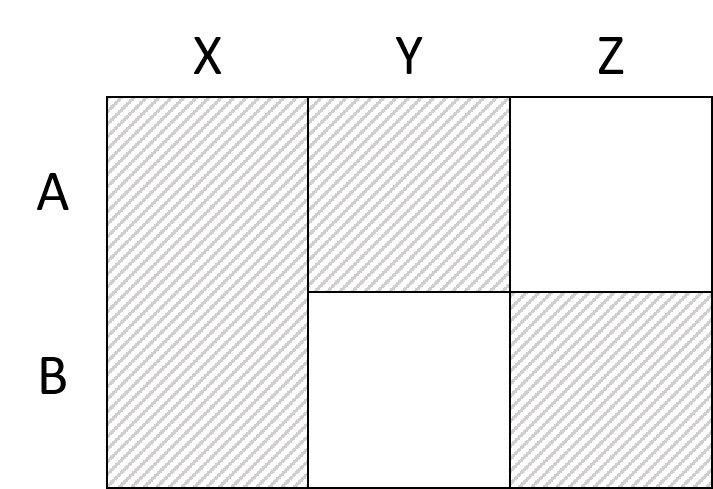}

	\caption{Missing Data Pattern of a Data Fusion Situation}
	\label{fig:fusSchem}
\end{figure}

Following the suggestion by \citet{Rubin.1986} to consider data fusion as file concatenation leads to the particular missing-by-design pattern \citep[see e.g.][ch. 4]{Rassler.2002}, and Figure \ref{fig:fusSchem} displays this schematic pattern if we stack two originally independent data sources A and B. The blank parts are missing and the corresponding variables were initially not part of the particular study, i.e. variables $Z$ are not observed in the first original study A (upper part of the stacked data set), and variables $Y$ are not observed in the second original study B (lower part of the stacked data set). In this respect, we denote variables which are observed in both studies as $X$ in the following, and we further denote variables relevant for the analysis which are only part of study A (but unobserved in study B) as $Y$ and, analogously, variables required for the analysis which are only observed in study B (but unobserved in study A) as $Z$.

A typical data fusion analysis objective is based on variables $Y$ and $Z$, and from the schematic overview it is apparent that we need identifying assumptions for the joint distribution of $f(Y,Z)$. In most imputation variants, either fully parametric or matching-based, an implicit \textit{Conditional Independence Assumption} (CIA) is made, which was first pointed out by  \citet{Sims.1972} in a comment on a technical report \citep{Okner.1972}. It states that any association between $Y$ and $Z$ is a function of $X$, i.e. $f(Y|X,Z)=f(Y|X)$ and, analogously, $f(Z|X,Y)=f(Z|X)$. This, for instance, yields a correlation of zero between $Y$ and $Z$ if conditioned on $X$. We will, however, not consider violations of distributional assumptions for $f(Y,Z)$ as part of this research. \citet{Rodgers.1984} extensively discussed the shortcomings of the CIA within a comprehensive simulation study, but in recent years several publications have addressed this issue by proposing to introduce auxiliary information (\citealp[see e.g.][]{Singh.1993};\citealp{Fosdick.2016}).

\subsection{Implementation in Practice}

Technically, we can apply any sophisticated method for handling missing data to this artificially created missing-by-design data situation, such as fully parametric multiple imputation or single imputation with variance correction, or Maximum Likelihood-based methods. The majority of empirical data fusions seem, however, to be based on some form of nearest neighbour matching \citep[see e.g.][ch. 2.4]{Koschnick.1995, DOrazio.2006}, and we assume that there are at least two reasons for it:
\begin{enumerate}
	\item Nearest neighbour based imputation methods are often more robust to model misspecification than fully parametric methods \citep[see e.g.][]{KollerMeinfelder.2009};
	\item The synonymous term 'Statistical Matching' already suggests matching-based methods to the practitioners as the most viable alternative.
\end{enumerate}

Although the missing-data pattern displayed in Figure \ref{fig:fusSchem} suggests that both missing parts could be imputed within the new stacked data set, it is far more common that only one of the original studies is used for data fusion analysis. Staying true to the matching concept, this study is labeled the \textit{recipient} study, whereas we refer to the study that 'donates' data from its observations as the \textit{donor} study \citep[see e.g.][]{Gabler.1997,vanderPutten.2002}. This applies to the content-based aspect of this paper, where EU-SILC represents the recipient study that has to be extended by the missing consumption expenditures, while HBS donates the household consumption information and, therefore, serves as the donor study \citep[see e.g.][]{Tonkin.2017}. This implies that, referring to the missing-by-design pattern displayed in Figure \ref{fig:fusSchem}, study A equals EU-SILC with the observed income variables $Y$  and study B corresponds to the HBS with the observed consumption variables $Z$. The aim is to expand the EU-SILC data file by the household consumption expenditures, i.e. imputing the missing $Z$ information in study A in order to provide a joint analysis of the income ($Y$) and consumption ($Z$) variables originally not jointly observed.

\subsection{Overview of Traditional Data Fusion Algorithms}

Traditionally, as already pointed out, data fusions are conducted using some form of covariate-based nearest neighbour matching methods \citep[see e.g.][]{Rodgers.1984,Koschnick.1995}. These algorithms match data on observations that are as close as possible with regard to their common $X$ characteristics, i.e. by imputing the missing $Z$ values in the recipient data file by the observed $Z$ values of the most similar donor observation according to the common $X$ variables \citep[see e.g.][]{vanderPutten.2002,Kiesl.2005}. Usually, either a certain distance metric is applied that accounts for the different scale levels of $X$, e.g. the distance proposed by \citet{Gower.1971}, or all $X$ variables will be categorised such that (alleged exact) matches can be identified in both data files \citep[ch. 2.4]{DOrazio.2006}. In the latter case, only zero distances between the $X$ variables are considered. 
In addition, however, there also exist fully parametric approaches. Such methods are based on regressions of $Z$ on $X$ within the donor file and subsequently estimate the missing $Z$ values in the recipient file by means of the computed regression parameters (see e.g. \citealp[ch. 2.2]{DOrazio.2006}; \citealp{Gilula.2006}). 

While coviarate-based methods are non-parametric as they are not subject to distributional assumptions, PMM can be considered as mixed (or semi-parametric) methods between coviarate-based algorithms and fully parametric approaches.
Note that Eurostat discusses different data fusion methods in previous working papers as well, and refers to semi-parametric algorithms as mixed methods between non-parametric and parametric approaches \citep{Leulescu.2013,Webber.2013,Tonkin.2017}. However, their 'mixed methods' are slightly different to the multivariate variant of PMM proposed by \citet{Little.1988}, as they consider ranks to identify suitable matches.

Nearest neighbour methods, in general, can be applied as either 'unconstrained' or 'constrained' matching \citep{Rodgers.1984,Rubin.1986}. Unconstrained matching allows a donor observation to be matched with a recipient record multiple times, whereas constrained matching limits the use of a donor observation to a predefined integer, for example to one. We focus on the unconstrained version for both, RHD and PMM, in the following, as there are many other sources discussing and comparing the two alternatives (\citealp[see e.g.][p. 53-63]{Rassler.2002}; \citealp{Kiesl.2005}).

\section{Random Hot Deck and Predictive Mean Matching}\label{sec:Algo}

In this section we will provide some details on the two methods, RHD and PMM, we want to compare within the subsequent Monte Carlo study, before we focus on describing the aforementioned differences of both methods in detail, and how these differences might affect the analysis results of our MC study.

\subsection{Random Hot Deck (RHD)}

In general, RHD randomly assigns observations from the donor file to observations of the recipient file. The missing $Z$ values for each recipient record are then imputed by the corresponding $Z$ values of its assigned donor observation. However, the random allocation between recipient and donor observations is usually carried out within homogeneous subgroups, for example only within the same gender category. Thus, in this example, female (male) donor observations can only be assigned to female (male) recipient observations \citep[ch. 2.4.1]{DOrazio.2006}.

We also apply RHD within homogeneous subgroups analogously to \citet{Eurostat.2018} and \citet{Lamarche.2018}. For any specific variable $Z_r$ (with $r=1,\ldots,p_{don}$) of $\mathbf{Z} = (Z_1,\ldots,Z_{p_{don}})$ stemming from the donor data file, the detailed fusion algorithm can be described as follows: First, in order to identify relevant matching classes that serve as homogeneous subgroups, all common variables $\mathbf{X}=(X_1,\ldots,X_p)$ that have a metric scale level are categorised. For example, age is transferred to rough age groups or income to income quintiles. Thus, all $\mathbf{X}$ variables are at most ordinal scaled and only zero distances between the recipient and the donor observations are allowed. Subsequently, a stepwise selection based on OLS regression of $Z_r$ on $\mathbf{X}$ in the donor file is implemented in order to select common $\mathbf{X}$ variables that have an acceptable explanatory power for $Z_r$. Along the stepwise-selected $X$ variables, an auxiliary variable is created that concatenates the respective values of $X$ for each observation in the recipient and the donor file. This results in a stratum characteristic for each donor and recipient record that form the homogeneous subgroups. The random assignment between the donor and the recipient observations is only conducted within the same stratum, i.e. every donor record is only permitted to be matched with a recipient record that has exactly the same (categorical) characteristics with respect to the stepwise-selected $X$ variables \citep{Eurostat.2018,Lamarche.2018}.

In order to ensure enough donor observations ($s_{l,don}$) compared to recipient records ($s_{l,rec}$) for each stratum level $l$, the following threshold is set: \begin{eqnarray*}
\frac{s_{l,rec}}{s_{l,don}}\quad\geq\quad c\ \cdot\ \frac{s_{rec}}{s_{don}}\ \text{,}
\end{eqnarray*}
where the constant $c$ is set as a rule of thumb to $c=3$ \citep[see][p. 13]{Lamarche.2018}. As long as 90\% of the sample do not exceed this threshold,\footnote{In the case of equal sample sizes of the recipient and the donor data file, the threshold means that a maximum of three recipients can be assigned to one donor -- for a maximum of 10 \% of the recipients, a deviation from this threshold is allowed.} the stepwise-selected $X$ variables are retained and the Random Hot Deck within each subgroup is performed. Otherwise, the process will be reiterated, with the maximum subset of the $\mathbf{X}$ variables to be selected by the stepwise selection\footnote{The maximum subset of $\mathbf{X}$ variables is controlled by the \texttt{nvmax} argument within the \texttt{regsubsets()} function from the \texttt{leaps} package \citep{Lumley.2020}.} being reduced by 1 for each iteration step. If there are still recipients left who cannot be assigned to a donor, a second round of allocation is conducted with $c=2$ and without a tolerance specification \citep{Eurostat.2018,Lamarche.2018}.

\subsection{Predictive Mean Matching (PMM)}
\label{subsec:PMM}

Predictive Mean Matching is not frequently discussed as a dedicated data fusion method, but it has become popular as an imputation method in general, and is the default method for metric-scale variables in the \texttt{R} package \texttt{mice} \citep{vanBuuren.2020}. The method was first introduced by \citet{Rubin.1986} and \citet{Little.1988} for the simultaneous imputation of continuous variables. The basic idea is that for each missing value its 'predictive mean' \citep[p. 291]{Little.1988} based on regression (e.g. OLS) is compared with the predictive means of all observed values, and the predictive mean among the observed values with minimum distance serves as donor record, and its actually observed values is imputed \citep{Rubin.1986,Little.1988}.

The PMM algorithm for any specific variable $Z_r$ ($r=1,\ldots,p_{don}$) of $\mathbf{Z}$ is as follows: First, as with RHD, relevant $\mathbf{X}$ variables are selected using a stepwise selection based on OLS regression of $Z_r$ on $\mathbf{X}$. In contrast to RHD, all $\mathbf{X}$ variables can remain on their original scale level, i.e. metric variables are not categorised \citep{Meinfelder.2015}. By means of the regression equation, which includes the stepwise-selected $X$ variables, the predictive mean is then calculated for each observation in the recipient and the donor file. The search for corresponding donor observations is now performed using the Mahalanobis distance function as proposed by \citet{Little.1988}: \begin{eqnarray*}\label{mah-dist}
D_{i,j}=(\hat{z}_i-\hat{z}_j)^TS^{-1}_{Z_r|\mathbf{X}}(\hat{z}_i-\hat{z}_j)
\end{eqnarray*}
$\text{with }i=1,\ldots,n_{rec}\text{ and } j=1,\ldots,n_{don}$,
where $\hat{z}_i$ corresponds to the predictive mean of the $i$-th observation from the recipient file and $\hat{z}_j$ corresponds to the predictive mean of the $j$-th observation from the donor file. $S^{-1}_{Z_r|\mathbf{X}}$ denotes the inverse variance-covariance matrix of the residuals from the regression of $Z_r$ on the stepwise selection subset of $\mathbf{X}$, by which the distance is weighted.

\subsection{Conceptual Differences of the two Algorithms}\label{subsec:AlgoDiffs}

Traditional co-variate-based nearest neighbour methods like Random Hot Deck assign per default equal weights to all $X$ variables, without taking into account the explanatory power or any other definition of relevance regarding the specific variables. 
Sometimes, as is the case with the considered implementation of the RHD algorithm, a regression-based stepwise selection of relevant $X$ variables precedes the distance computation. While the $X$ variables selected via the stepwise algorithm might be adequate predictors for the $Y$ and $Z$ variables to be matched, their explanatory power is unlikely to be equally high. Technically, any variable selection to identify 'suitable' $X$ variables equates a weighting process, where weights for $X$ variables are either $0$ or $1$. Although covariate-based methods like Random Hot Deck are widely used, the issue of unequal explanatory power of the common $X$ variables results in an optimization problem for identifying the 'best' matches. PMM, on the other hand, addresses this optimization problem by weighting the distance computation with $S^{-1}_{Z_r|\mathbf{X}}$, i.e. the variance-covariance matrix of the residuals from $Z_r|\mathbf{X}$.  Therefore, the main difference of PMM compared to RHD is in the distance processing: Within RHD, every stepwise-selected $X$ variable has the same weight in finding those 'best' matches within the recipient and donor files. The reasonable possibility that there exist selected $X$ variables that have a better explanatory power for $Z_r$ than other common variables is ignored by RHD. PMM takes the unequal explanatory power of the selected $X$ variables into account, as the predictive means depend on the estimated regression parameters. Additionally, distances between predictive means are weighted with the inverse of the residual covariance matrix from the regressions of $Z_1,\ldots,Z_{p_{don}}$ on $\mathbf{X}$. This means that the better some $Z_r$ variable can be explained by the covariates $\mathbf{X}$, the more distances between predictive means of potential donors and recipients are penalized. This may indicate a more sophisticated distance processing in favour of PMM and, consequently, we expect PMM to result in a better fusion performance than the RHD method. Therefore we formulate the following hypothesis: \textit{Since Predictive Mean Matching provides a more precise distance processing, PMM leads to a better fusion result than RHD.} The upcoming simulation study is designed to provide a differentiated and detailed scrutiny of this underlying hypothesis.

\section{Simulation Design}
\label{sec:Sim_Design}
Since the motivation for the present analysis is derived from the findings published by Eurostat \citep{Webber.2013,Tonkin.2017,Lamarche.2018}, our aim for the data basis of the simulation study is to stay as close as possible to the relevant official statistics data sources.

With respect to the analysis objective of our comparison between RHD and PMM, we deviate from the focus of the previous studies, where emphasis was mainly put on preserving marginal distributions of the donor study within the fused data source. Instead, we concentrate on bivariate associations between common variables $X$ and fused variables $Z$ as well as on the primary objective of any data fusion, the joint distribution of the not jointly observed variables $Y$ and $Z$. As stated above, we expect PMM to preserve the correlations between the variables $Y$ and $Z$ as well as between $X$ and $Z$ better than RHD, since correlations suffer more from non-exact matches than marginal distributions. Therefore, our simulation study focuses on evaluating the performance of both algorithms with respect to associations among different variables.

\subsection{Database}\label{subsec:Datenbasis}

We conduct a Monte Carlo (MC) study based on Scientific Usefiles (SUFs) of EU-SILC from 2015.\footnote{Eurostat and the NSIs also focus on matching of the EU-SILC and HBS data files from 2015.} In order to ensure a sufficiently large surrogate population, we combine EU-SILC data for Germany ($N_{DE}=12,533$) and France ($N_{FR}=10,855$). This leads to a total number of $N=23,418$ observations from which we draw $k=1,000$ random samples that we subsequently split into two data files which serve as substitutes for EU-SILC as recipient and HBS as donor file. 
For our simulation purposes, all data are based on EU-SILC due to the necessity of knowing the 'true' joint distribution of the simulated $Y$ and $Z$ variables and their correlations.

As common variables we select seven $X$ variables
that represent those variables Eurostat had selected for their analyses \citep{Leulescu.2013,Lamarche.2018}.
Table \ref{tab_Xvar} shows an overview of the $p=7$ variables $X_1,\ldots,X_7$ used in the upcoming simulation study as well as the respective value range and measurement level. It can be seen that for RHD we have to categorise all variables, whereas for PMM the variables age ($X_2$) and income ($X_7$) keep their original metric scale level. 

\begin{table}[H]

\begin{center}
\small
\caption{Overview of the Common $X$ Variables}
    \label{tab_Xvar}
\begin{tabular}{l|c|c}
\hline \hline

\multirow{2}{*}{\textbf{Common $\boldsymbol{X}$ Variables}}  & \multicolumn{2}{c}{\textbf{Range / Measurement Level}} 
\\ \cline{2-3}
& \textbf{RHD} & \textbf{PMM} \\ \hline
$X_1$: Activity Status of RP$\,^{a}$ & 1 to 5 / categorical  & 1 to 5 / categorical  
\\
$X_2$: Age of RP$\,^a$& 1 to 8 / categorical  &
acc. $X_2$ / metric \\
$X_3$: Population Density Level & 1 to 3 / categorical  & 1 to 3 / categorical \\
$X_4$: Dwelling Type$\,^{b}$ & 1 to 4 / categorical  & 1 to 4 / categorical 
 \\
$X_5$: Tenure Status & 1 to 5 / categorical  & 1 to 5 / categorical  
 \\
 $X_6$: Main Source of Income$\,^c$ & 1 to 2 / categorical & 1 to 2 / categorical \\
$X_7$: Income & 1 to 5 / categorical  & acc. $X_7$ / metric   \\

    \hline \hline

        \multicolumn{3}{l}{\begin{footnotesize}$^{a}\,$RP: 'Reference person' (interviewed person of the household);\end{footnotesize}}\vspace{-0.3cm}\\
         \multicolumn{3}{l}{\begin{footnotesize}$^{b}\,$Actual range 1 to 5, category 5 is empty.\end{footnotesize}}\vspace{-0.3cm}\\
         \multicolumn{3}{l}{\begin{footnotesize}$^{c}\,$Here, the missing values also form a category (coded as 9); 
        \end{footnotesize}}\vspace{-0.1cm}\\
         \multicolumn{3}{l}{\begin{footnotesize}Source: EU-SILC 2015 SUF: DE, FR.\footnotemark\end{footnotesize}}
        
\end{tabular}

\end{center}
\end{table}\vspace{-0.8cm}
\footnotetext{EU-SILC 2015 SUF DE: European Union Statistics on Income and Living Conditions, 2015.
Scientific Usefile Germany.

EU-SILC 2015 SUF FR: European Union Statistics on Income and Living Conditions, 2015.
Scientific Usefile France.}

The variable \textit{activity status} ($X_1$) contains information about the types of employment (self-employed or non-self-employed, pensioner, unemployed, etc.)  \citep[p. 285]{Eurostat.2016}. Details concerning the generation and recoding of this variable can be found in the Appendix. The \textit{population density level} ($X_3$) indicates the population density of the current residential area. For the German EU-SILC SUF, this variable is empty (probably due to confidentiality reasons). Therefore, we generated the respective values randomly for German data but under consideration of the univariate distribution of the original 2015 EU-SILC data for Germany \citep[see][]{Eurostat.2020}. The generation is documented in the Appendix. \textit{Dwelling type} \citep[p. 173]{Eurostat.2016} reflects the type of accommodation (residential building, flat, etc.) \citep[p. 173]{Eurostat.2016}.
The \textit{tenure status} ($X_5$) combines information about the ownership status of the housing unit (sole owner, tenant, etc.) and on the (classified) rental costs incurred in case of a rental contract \citep[p. 174, 181]{Eurostat.2016}. 
The binary variable $\textit{main source of income}$ ($X_6$) distinguishes between (1) income from self-employment or non-self-employment, property, ownership and assets and (2) income from pensions, social benefits and other transfers (\citealp[p. 20, 27-28]{Eurostat.2013}; \citealp[p. 7, 313-316, 322-336]{Eurostat.2016}).
Its generation and recoding is documented in the Appendix. \textit{Income} reflects the 'total disposable household income' \citep[p. 209]{Eurostat.2016} and, for RHD, is recoded into five income quintiles.

For the specific variables $Y$ and $Z$, which represent the income variables of EU-SILC ($Y$) and the consumption variables of HBS ($Z$) that are actually not jointly observed, we select $p_{silc}=p_{hbs}=2$ substitutes each, i.e. $Y=(Y_1,Y_2)$ and $Z=(Z_1,Z_2)$, from the database. This underlines the possibility that the univariate data fusion (imputing only \textit{one} HBS variable) can also be performed in a multivariate setting with more than one specific HBS characteristic. It becomes obvious that an exact coverage of the specific variables $Y$ and $Z$ is only possible for the income variables from EU-SILC, because the database itself consists of EU-SILC. However, since both, statistical and methodological conclusions, are of interest, it is more important to ensure the same measurement level for the respective income and expenditure substitutes and, therefore, it is essential to select \textit{metric} variables.

For $Y_1$ we choose the variable 'total disposable household income before social transfers including old-age and survivor's benefits' \citep[p. 209]{Eurostat.2016} and for $Y_2$ the variable 'interest, dividends, profit from capital investments in unincorporated business' \citep[p. 214]{Eurostat.2016}. The variables 'total household gross income' \citep[p. 207]{Eurostat.2016} and 'total disposable household income before social transfers other than old-age and survivor's benefits' \citep[p. 209]{Eurostat.2016} are selected for $Z_1$ and $Z_2$. Table \ref{tab_YZvar} displays an overview of the specific $Y$ and $Z$ variables used for the simulation study and the corresponding measurement level.

\begin{table}[H]

\begin{center}
\small
\caption{Overview of the Specific Substitute Variables for EU-SILC ($Y$) and HBS ($Z$)} 
    \label{tab_YZvar}
\begin{tabular}{l|l}
\hline \hline

\textbf{Specific EU-SILC Substitute Variables ($\boldsymbol{Y}$)} & \textbf{Measurement Level} \\ \hline
$Y_1$: Total disposable household income before social & metric\vspace{-0.3cm} \\ \phantom{$Y_1$: }transfers including old-age and survivor's benefits & \\ \hline
$Y_2$: Interest, dividends, profit from capital investments in & metric \vspace{-0.3cm}\\
\phantom{$Y_2$: }unincorporated business & \\ \vspace{-0.4cm}\\ \hline\hline

\textbf{Specific HBS Substitute Variables ($\boldsymbol{Z}$)} & \textbf{Measurement Level} \\ \hline
$Z_1$: Total household gross income & metric\\ \hline
$Z_2$: Total disposable household income before social & metric\vspace{-0.3cm}\\
\phantom{$Z_2$: }transfers other than old-age and survivor's benefits & 

\\

    \hline \hline

      \multicolumn{2}{l}{\begin{footnotesize}Source: EU-SILC 2015 SUF: DE, FR.\end{footnotesize}}
        
\end{tabular}

\end{center}
\end{table}\vspace{-0.8cm}

\subsection{Monte Carlo Study}
\label{subsec:MCmeth}

Our MC study is structured as follows: First, we draw $k=1,000$ random samples without replacement (Jackknife) from the data specified above. Subsequently, for each random draw we generate the specific missing data pattern underlying data fusion scenarios (see Fig. \ref{fig:fusSchem}) and impute the missing $Z_1$ and $Z_2$ values in the $k=1,000$ simulated data files via 
RHD on the one hand and PMM on the other hand. 

More specifically, each random draw leads to a simulated data file that represents EU-SILC with the observed variables $X=(X_1,X_2,X_3,X_4,X_5,X_6,\newline X_7)$ and $Y=(Y_1,Y_2)$ without information on the $Z$ variables as well as to a simulated data file that represents HBS with the observed variables $X=(X_1,X_2,X_3,X_4,X_5,X_6,X_7)$ and $Z=(Z_1,Z_2)$ that in turn contains no information about the $Y$ variables. 'Stacking' both data sources results in the specific missing data pattern displayed in Figure \ref{fig:fusSchem}. We impute the missing $Z_1$ and $Z_2$ values in the simulated EU-SILC data file using the two proposed data fusion algorithms, RHD and PMM. Thus, the imputed $Z$ values in the matched data file reflect an artificial distribution $\widetilde{Z}=(\widetilde{Z}_1,\widetilde{Z}_2)$. After the imputation step the correlations between $Y$ and $\widetilde{Z}$ as well as between the metric $X$ variables ($X_2$: age and $X_7$: income) and $\widetilde{Z}$ are then calculated and compared to the true correlations known from the surrogate population with $N=23,418$ individuals as described in Section \ref{subsec:Datenbasis}. We apply single imputation ($M=1$), resulting in point estimates for the correlations. This process, from sampling to imputation to the computation of correlations is performed with $k=1,000$ simulation draws.

In order to get a rough understanding of how sensitive the performance of RHD and PMM could be with regard to the sample sizes of the recipient data file ($n_{silc}$) and the donor data file ($n_{hbs}$), we vary the sample size $n$. Of particular interest is the extent to which an excessive number of donors ($n_{silc}<<n_{hbs}$) compared to an equal recipient and donor ratio ($n_{silc}=n_{hbs}$) has an effect on the performance of both data fusion algorithms. Therefore, the MC simulation is performed twice using different sample sizes $n_1$ and $n_2$. For both simulation scenarios we consider $n_{1_{silc}}=n_{2_{silc}}=400$ observations for EU-SILC. In the first simulation scenario we also assign $n_{1_{hbs}}=400$ units to the HBS. However, in the second scenario we choose a significantly higher number of donor observations, namely $n_{2_{hbs}}=3,600$ for the HBS data. This leads to a sample size of $n_1=800$ with $n_{1_{silc}} = n_{1_{hbs}} = 400$ for the first scenario as well as to a sample size of $n_2=4,000$ with $n_{2_{silc}} = 400$ and $n_{2_{hbs}} = 3,600$ for the second scenario.

The MC simulation is conducted using \texttt{R} \citep{RCoreTeam.2019}, and we use packages \texttt{StatMatch} \citep{DOrazio.2019} and \texttt{BaBooN} \citep{Meinfelder.2015} for RHD and PMM, respectively.

\section{Results}\label{sec:results}

Since we start out with complete samples, where parts of the data are removed to mimic a data fusions scenario, we know the true parameter values even for those parameters pertaining to $f(Y,Z)$, but the fusion algorithms implicitly rely on the CIA, and the theoretically correct values under this assumption are displayed as additional benchmarks in the results. Note, however, that while data fusion requires assumptions regarding the joint distribution of $Y$ and $Z$, the identification problem and the natural uncertainty arising from it are not the primary focus of our work, but has been covered by many other authors \citep[see e.g.][]{Kiesl.2006,Conti.2012,Fosdick.2016,Endres.2019}.

\subsection{Correlations Between $\boldsymbol{Y}$ and $\boldsymbol{\widetilde{Z}}$}\label{subsec:results_YZ}

As stated in Section \ref{subsec:AlgoDiffs}, we expect PMM to outperform RHD for bivariate associations. Hence, PMM should be able to reproduce the unobserved correlations between $Y$ and $Z$ more accurately than RHD. Table \ref{tab_true_val_YZ} displays the correlations between the specific variables $Y=(Y_1,Y_2)$ and $Z=(Z_1,Z_2)$, resulting from the artificial population consisting of $N=23,418$ observations. These population correlations are used as true parameters for graphical diagnostics throughout this section and for Bias and MSE. The correlations between $Y_1$ and $Z_1$ as well as the correlations between $Y_1$ and $Z_2$ (0.87 and 0.85) are relatively high, whereas for $Y_2$ and $Z_1$ as well as for $Y_2$ and $Z_2$ we observe moderate correlations (0.42 and 0.47).

\begin{table}[H]

\begin{center}
\small
\caption{True Parameters for $\rho_{YZ}$}
    \label{tab_true_val_YZ}
    
\begin{tabular}{c|c|c|c}

\hline \hline

$\cor(Y_1,Z_1)$ & $\cor(Y_1,Z_2)$ & $\cor(Y_2,Z_1)$ &$\cor(Y_2,Z_2)$  \\ \hline 0.8665 & 0.8515 & 0.4211 & 0.4734 \\

    \hline \hline
      \multicolumn{4}{l}{\begin{footnotesize}Source: EU-SILC 2015 SUF: DE, FR.\end{footnotesize}}

\end{tabular}

\end{center}
\end{table}\vspace{-0.8cm}

\begin{figure}[H]

\begin{center}

    \includegraphics[width=\textwidth,height=0.7\textheight]{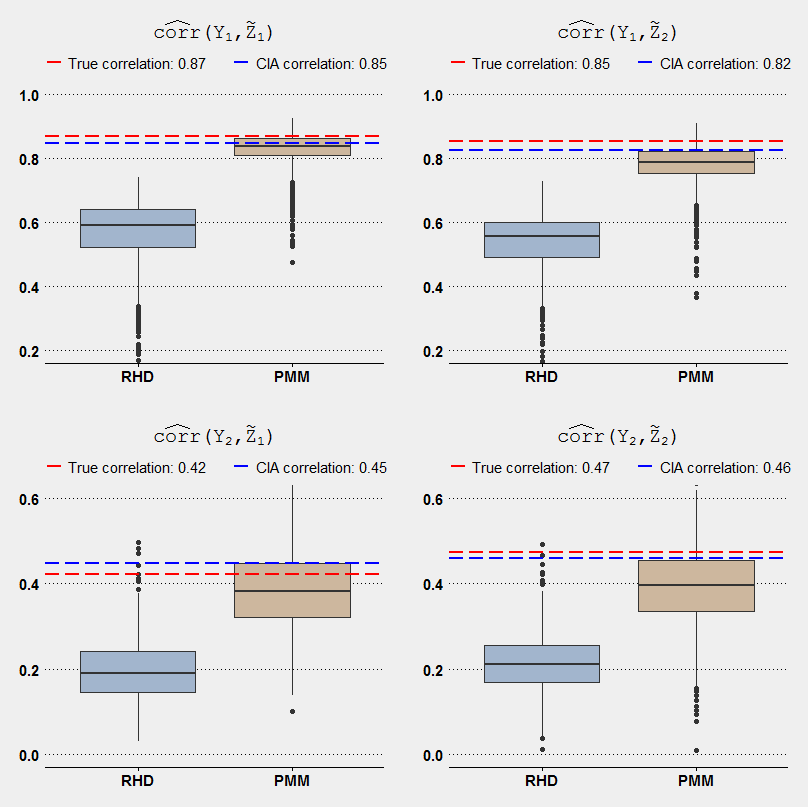}\\
    \begin{footnotesize}Source: EU-SILC 2015 SUF: DE, FR.\end{footnotesize}\vspace{-0.6cm}

    \caption{Boxplots --  MC distributions for $\widehat{\rho}_{Y\widetilde{Z}}$ with $n_1$}
    \label{Boxplot_YZ_n1}

    \end{center}
\end{figure}

Figure \ref{Boxplot_YZ_n1} displays the MC distributions of the estimated correlations over all $k=1,000$ MC simulation draws with equal number of recipients and donors ($n_{silc}=n_{hbs}=400$). For high original correlations of 0.87 and 0.85, we find convincing evidence that PMM is able to reproduce the true parameter values more accurately than RHD. While RHD never covers the immediate area around the true correlations of 0.87 and 0.85 -- the respective maximum for RHD amounts to 0.74 for $\caa$ and 0.73 for $\cab$ -- the distributions of the estimated correlations resulting from PMM for $\caa$ and $\cab$ are considerably close to the original parameter values. This also becomes clear by looking at the respective means: PMM produces mean correlations for $\caa$ of 0.83 and for $\cab$ of 0.78 and thus comes on average very close to the original values of 0.87 and 0.85, respectively. RHD generates mean correlations of 0.57 for $\caa$ and 0.54 for $\cab$ that deviate more strongly from the observed original parameter values. Furthermore, PMM reproduces the correlation between $Y_1$ and $Z_1$ more accurately than the respective relationship between $Y_1$ and $Z_2$. For moderate original correlations between $Y$ and $Z$ of 0.42 and 0.47 it can be seen that the superior performance in favour of PMM is slightly lower but still present. The MC distributions over all $k=1,000$ simulation draws illustrated in Figure \ref{Boxplot_YZ_n1} show for $\cba$ and $\cbb$ that the estimated PMM correlations cover the area around the true parameters more frequently while RHD tends to underestimate them. Consequently, the mean of the estimated correlations for $\cba$ and $\cbb$ resulting from RHD are negatively biased with 0.20 and 0.21, respectively, while PMM produces almost unbiased estimators with MC mean correlations of 0.38 and 0.40.

\vspace{-0.8cm}

\begin{figure}[H]
\begin{center}  
    \includegraphics[width=\textwidth,height=0.7\textheight]{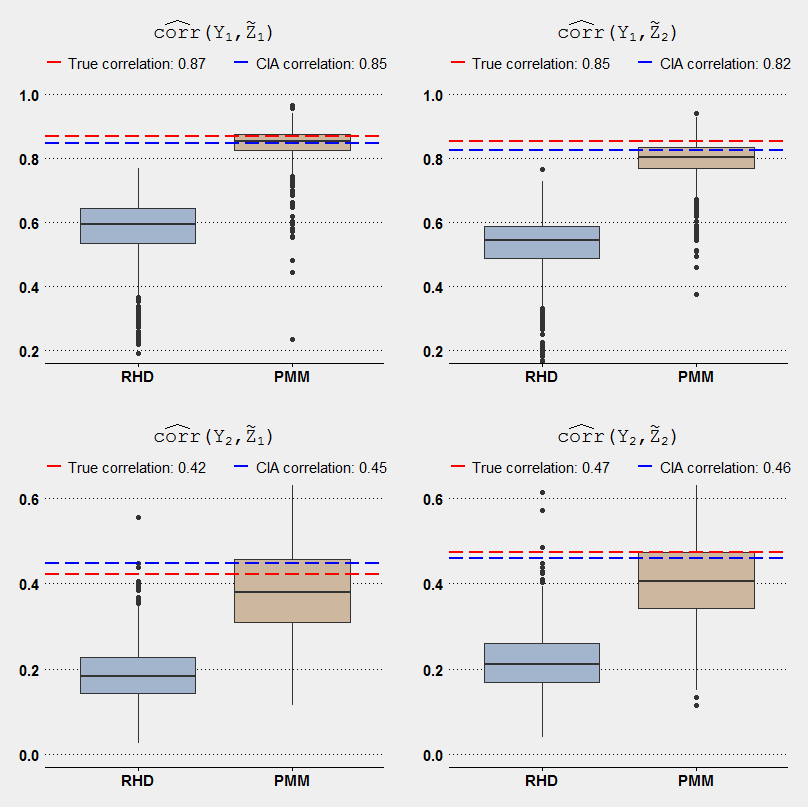}\\
    \begin{footnotesize}Source: EU-SILC 2015 SUF: DE, FR.\end{footnotesize}\vspace{-0.6cm} 
    
    \caption{Boxplots --  MC distributions for $\widehat{\rho}_{Y\widetilde{Z}}$ with $n_2$}
    \label{Boxplot_YZ_n2}
\end{center}
\end{figure}\vspace{-0.8cm}

The second scenario of the MC study contains an excessive number of donors in order to investigate mitigating effects on the results, if the proposed matching methods can choose from a larger donor pool, i.e. the overall sample size $n_2 = 4,000$ consists of $n_{hbs}=3,600$ donors versus $n_{silc}=400$ recipients. In general, no substantial change can be observed for the RHD results as we can see in the respective MC distributions 
illustrated in Figure \ref{Boxplot_YZ_n2}. The correlations resulting from PMM with $n_2$ indicate, compared to the PMM correlations with $n_1$ in the first scenario, in terms of $\caa$ and $\cab$ slightly improved results, since the bulk of MC distribution of the PMM correlations is even closer to the true values of 0.87 and 0.85. Accordingly, the mean correlations computed with PMM under $n_2$ increase marginally to 0.84 for $\caa$ and 0.80 for $\cab$. In contrast, with respect to $\cba$ and $\cbb$ a slightly smaller number of the $k=1,000$ PMM correlation estimates covers the area around the moderate true correlations of 0.42 and 0.47, while again the mean values marginally increase to 0.39 and 0.41 due to a higher outlier rate towards 1 that goes along with a somewhat higher variance. However, it should be noted that only small changes between $n_1$ and $n_2$ can be observed that should be treated with caution due to random fluctuations.

In addition to the true correlations of $Y$ and $Z$, we marked the respective correlations under CIA in Figures \ref{Boxplot_YZ_n1} and \ref{Boxplot_YZ_n2}, i.e. the theoretical correlations of $Y$ and $Z$, assuming independence if conditioned on $X$. This scenario underpins the relative advantage of PMM over RHD, as the PMM correlations are close to the correct values, whereas RHD fails to accurately reproduce the correlation structure of $Y$ and $Z$, even though correlations under CIA are in this case close to the true correlations.

Consequently, the PMM correlation estimates have a smaller bias compared to the RHD estimates, as displayed in Table \ref{tab_BIAS_YZ}. While PMM yielded more outlier results within the MC simulation study, the MSE for PMM is still much lower than RHD's with respect to all four correlations of interest over both scenarios (see Tab. \ref{tab_MSE_YZ}).

\begin{table}[H]

\begin{center}
\small
\caption{Bias of $\widehat{{\rho}}_{Y\widetilde{Z}}$}
    \label{tab_BIAS_YZ}
\begin{tabular}{l|l|c|c|c|c}

\hline \hline

&& $\widehat{\cor}(Y_1,\widetilde{Z}_1)$ & $\widehat{\cor}(Y_1,\widetilde{Z}_2)$ & $\widehat{\cor}(Y_2,\widetilde{Z}_1)$ &$\widehat{\cor}(Y_2,\widetilde{Z}_2)$  \\ \hline
\multirow{2}{*}{$n_1$} &\textbf{RHD} & 0.2973 & 0.3163 & 0.2250 & 0.2606 \\ &\textbf{PMM} & 0.0405 & 0.0731 & 0.0367 & 0.0773 \\ \hline\hline
\multirow{2}{*}{$n_2$}& \textbf{RHD} & 0.2878 & 0.3224 & 0.2332 & 0.2583 \\ & \textbf{PMM} & 0.0245 & 0.0563 & 0.0319 & 0.0632 \\ \hline\hline
      \multicolumn{5}{l}{\begin{footnotesize}Source: EU-SILC 2015 SUF: DE, FR.\end{footnotesize}}
\end{tabular}

\end{center}
\end{table}\vspace{-0.8cm}

\begin{table}[H]

\begin{center}
\small
\caption{MSE of $\widehat{{\rho}}_{Y\widetilde{Z}}$}
    \label{tab_MSE_YZ}
\begin{tabular}{l|l|c|c|c|c}

\hline \hline

&& $\widehat{\cor}(Y_1,\widetilde{Z}_1)$ & $\widehat{\cor}(Y_1,\widetilde{Z}_2)$ & $\widehat{\cor}(Y_2,\widetilde{Z}_1)$ &$\widehat{\cor}(Y_2,\widetilde{Z}_2)$  \\ \hline
\multirow{2}{*}{$n_1$}&\textbf{RHD} & 0.0987 & 0.1089 & 0.0557 & 0.0728 \\ &\textbf{PMM} & 0.0050 & 0.0096 & 0.0109 & 0.0152 \\ \hline\hline \multirow{2}{*}{$n_2$} & \textbf{RHD} & 0.0913 & 0.1113 & 0.0587 & 0.0719 \\ &\textbf{PMM} & 0.0040 & 0.0071 & 0.0133 & 0.0147 \\ \hline\hline
      \multicolumn{5}{l}{\begin{footnotesize}Source: EU-SILC 2015 SUF: DE, FR.\end{footnotesize}}

\end{tabular}

\end{center}
\end{table}\vspace{-0.8cm}

\subsection{Correlations Between $\boldsymbol{X}$ and $\boldsymbol{\widetilde{Z}}$}\label{subsec:results_XZ}

Apart from the reproduction of the joint information between the variables not jointly observed, the preservation of the distribution between the common variables $X$ and the specific variables $Z$ can be regarded as a minimum requirement, as it does not rely on any identifying assumptions. Table \ref{tab_true_val_XZ} shows the true correlations $\rho_{XZ}$ resulting from the data base that represents our artificial population specified in Section \ref{subsec:Datenbasis}. For $X$ we consider the metric variables $X_2$ and $X_7$. 
Here, correlations between $X_2$ and $Z_1$ as well as between $X_2$ and $Z_2$ are relatively low with $-0.13$ and $-0.04$, respectively, while correlations for $X_7$ and $Z_1$ as well as for $X_7$ and $Z_2$ are rather high with 0.97 each. 

\begin{table}[H]

\begin{center}
\small
\caption{True Parameters for $\rho_{XZ}$}
    \label{tab_true_val_XZ}
\begin{tabular}{c|c|c|c}

\hline \hline

$\cor(X_2,Z_1)$ & $\cor(X_2,Z_2)$ & $\cor(X_7,Z_1)$ &$\cor(X_7,Z_2)$  \\ \hline $-0.1314$ & $-0.0413$ & 0.9695 & 0.9735 \\

    \hline \hline
      \multicolumn{4}{l}{\begin{footnotesize}Source: EU-SILC 2015 SUF: DE, FR.\end{footnotesize}}

\end{tabular}

\end{center}
\end{table}\vspace{-0.8cm}

\begin{figure}[H]
\begin{center}

    \includegraphics[width=\textwidth,height=0.7\textheight]{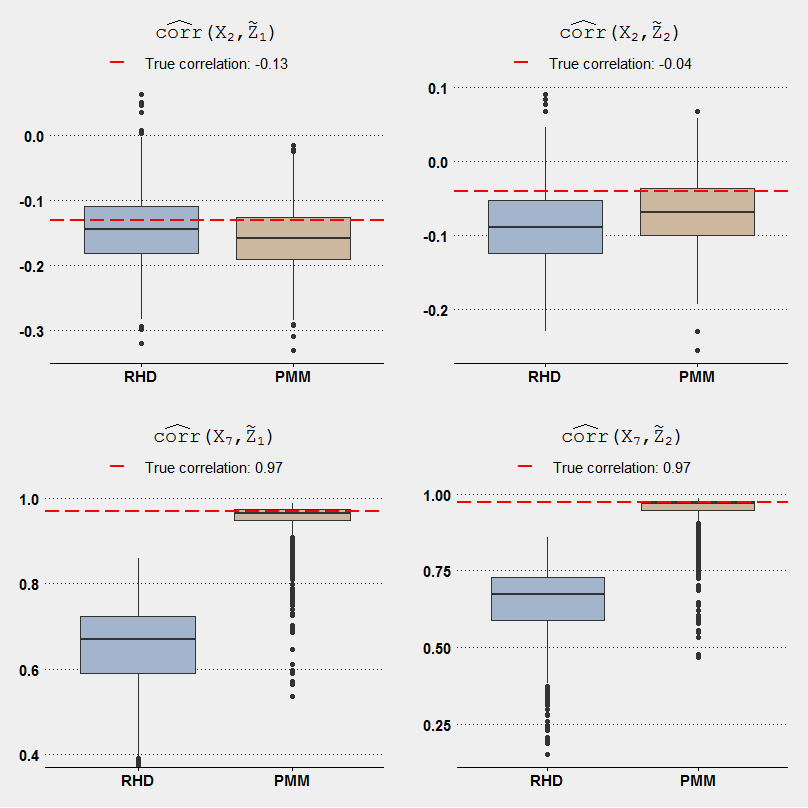}\\
    \begin{footnotesize}Source: EU-SILC 2015 SUF: DE, FR.\end{footnotesize}\vspace{-0.6cm}
    \caption{Boxplots -- MC distributions for $\widehat{\rho}_{X\widetilde{Z}}$ with $n_1$}
    \label{Boxplot_XZ_n1}
\end{center}
\end{figure}

\begin{figure}[H]
\begin{center}

    \includegraphics[width=\textwidth,height=0.7\textheight]{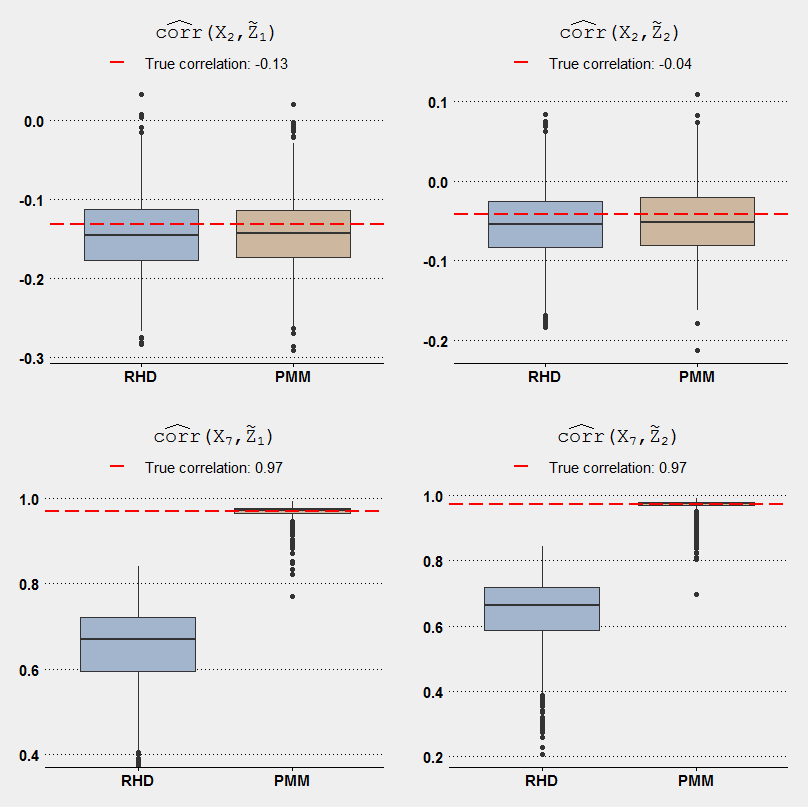}\\
    \begin{footnotesize}Source: EU-SILC 2015 SUF: DE, FR.\end{footnotesize}\vspace{-0.6cm}
    \caption{Boxplots -- MC distributions for $\widehat{\rho}_{X\widetilde{Z}}$ with $n_2$}
    \label{Boxplot_XZ_n2}
\end{center}
\end{figure}

As can be seen in the respective MC distributions displayed in Figure \ref{Boxplot_XZ_n1}, both methods struggle with preserving the low correlations between $X_2$ and both specific variables. One possible explanation is that variable $X_2$ was not included in the backward-deletion selected matching model, as variable $X_7$ explains variables $Z_1$ and $Z_2$ almost perfectly, thus accidentally creating an uncongeniality issue \citep{Meng.1994}.
For the scenario with excessive donors this phenomenon vanishes which can be seen in Figure \ref{Boxplot_XZ_n2}, because the larger sample size increased the probability of $X_2$ to remain in the underlying model. 

As expected, PMM does once again much better in preserving high correlations, as the respective correlation estimates for $\xba$ and $\xbb$ come very close to the true parameter values and amount on average to 0.95 under $n_1$ and approximately to the real parameter value of 0.97 under $n_2$. RHD produces mean correlations of 0.64 ($n_1$) and 0.65 ($n_2$) for $\xba$ as well as 0.65 ($n_1$) and 0.64 ($n_2$) for $\xbb$, respectively, that fall far behind the real observed parameter values. Investigating the bias and the MSE underlines these findings in favour of PMM, as displayed in Tables \ref{tab_BIAS_XZ} and \ref{tab_MSE_XZ}.

\vspace{-0.8cm}

\vspace{-0.8cm}

\begin{table}[H]

\begin{center}
\small
\caption{Bias of $\widehat{{\rho}}_{X\widetilde{Z}}$}
    \label{tab_BIAS_XZ}
\begin{tabular}{l|l|c|c|c|c}

\hline \hline

&& $\widehat{\cor}(X_2,\widetilde{Z}_1)$ & $\widehat{\cor}(X_2,\widetilde{Z}_2)$ & $\widehat{\cor}(X_7,\widetilde{Z}_1)$ &$\widehat{\cor}(X_7,\widetilde{Z}_2)$  \\ \hline \multirow{2}{*}{$n_1$} &
\textbf{RHD} & 0.0124 & 0.0474 & 0.3263 & 0.3263 \\ & \textbf{PMM} & 0.0264 & 0.0275 & 0.0224 & 0.0284 \\ \hline\hline \multirow{2}{*}{$n_2$} & \textbf{RHD} & 0.0138 & 0.0132 & 0.3235 & 0.3313 \\ & \textbf{PMM} & 0.0111 & 0.0103 & 0.0022 & 0.0042 \\ \hline \hline
      \multicolumn{5}{l}{\begin{footnotesize}Source: EU-SILC 2015 SUF: DE, FR.\end{footnotesize}} 
\end{tabular}

\end{center}
\end{table}\vspace{-0.8cm}

\begin{table}[H]

\begin{center}
\small
\caption{MSE of $\widehat{{\rho}}_{X\widetilde{Z}}$}
    \label{tab_MSE_XZ}
\begin{tabular}{l|l|c|c|c|c}

\hline \hline

&& $\widehat{\cor}(X_2,\widetilde{Z}_1)$ & $\widehat{\cor}(X_2,\widetilde{Z}_2)$ & $\widehat{\cor}(X_7,\widetilde{Z}_1)$ &$\widehat{\cor}(X_7,\widetilde{Z}_2)$  \\ \hline
\multirow{2}{*}{$n_1$}& \textbf{RHD} & 0.0032 & 0.0049 & 0.1197 & 0.1192 \\ & \textbf{PMM} & 0.0030 & 0.0029 & 0.0033 & 0.0051 \\ \hline\hline \multirow{2}{*}{$n_2$} & \textbf{RHD} & 0.0025 & 0.0022 & 0.1158 & 0.1210 \\ & \textbf{PMM} & 0.0022 & 0.0021 & 0.0004 & 0.0007 \\ \hline \hline
      \multicolumn{5}{l}{\begin{footnotesize}Source: EU-SILC 2015 SUF: DE, FR.\end{footnotesize}} 

\end{tabular}

\end{center}
\end{table}\vspace{-0.8cm}

\section{Conclusion}\label{sec:Conclusion}

The objective of our research was to compare two types of data fusion methods, Random Hot Deck as a representative of classical nearest neighbour hot deck methods as well as the alternative Predictive Mean Matching algorithm. Covariate-based variants like RHD are wide-spread for data fusion in practice, whereas Predictive Mean Matching is a popular method for conditional univariate sequential regression imputation algorithms \citep{vanBuuren.2011}, but is far less common as a data fusion alternative. In general, PMM can be classified as a semi-parametric method \citep{Rubin.1987}, since it requires an underlying parametric model for predicting the (conditional) means of missing and observed cases, which distinguishes it from the purely covariate-based statistical matching methods. This can be perceived as a drawback, as the 'non-parametric' covariate-based methods do not require this step. However, we still make implicit assumptions for covariate-based algorithms about the association between the common and the specific variables by deciding to use a particular distance metric. Since our goal is the joint analysis of these variables, we assume a particular data generating process (including identifying assumptions), and we implicitly assume the imputation method to be \textit{congenial} to the analysis model, i.e. associations among variables which are part of the analysis have to be accounted for by the imputation model as well \citep[for details see][]{Meng.1994, Xie.2017}. To some extent this drawback therefore can be viewed as advantageous, as PMM requires us to think about the nature of the relationship between the common $X$ variables and the specific $Z$ variables to be fused.
 
The already mentioned benefit of PMM is the implicit weighting of the common $X$ variables with respect to the specific $Z$ variables. Since exact matches are rare in purely discrete settings and impossible in continuous settings, distances between recipients and potential donors play a crucial role. And some $X$ variables usually turn out to be more relevant for the distance processing in order to find the 'nearest' donor observation and, therefore, we have to account for the unequal explanatory power of the common $X$ variables with regard to the specific $Z$ variables to be matched. Classical covariate-based methods assign weights to the $X$ variables, and these are either 1 or 0, depending on whether a particular variable $X_j$ is included in the distance measure. There is no straightforward procedure that decides which variables should be included, but, using \textit{all} potential common variables can lead to very inefficient matches. PMM, on the other hand, reverses the weighting to the specific variables using the Mahalanobis distance of the residuals. As stated in Section \ref{sec:Algo} in order to keep our research close to applied problems, we additionally included a variable selection scheme via backward-deletion for both algorithms to reduce the potentially high number of common variables to a more sensible subset of matching covariates. However, this does not invalidate the general perspective on the differences between PMM and RHD.

Moreover, adequate distance processing is even more important if the number of $X$ variables included in the data fusion process is relatively high. Additionally, the lower the number of potential donors, the higher, on average, the distances between recipients and matched donors \citep{Andridge.2010}.  Subsequently, the probability for exact matches decreases and the way we are dealing with non-exact matches becomes more relevant.

The RHD algorithm as implemented by Eurostat requires a categorisation of metric variables which may cause a loss of information. The effect is similar to the above mentioned problem of increasing distances between recipients and matched donors, as the categorisation might lead to matches which are not the best possible ones under the original metric scale. Findings from additional simulations using Gower distance (without categorisation) suggest, however, that PMM is still superior to covariate-based matching, although the gap becomes slightly closer in the excessive donor scenario.

We did not consider constrained matching \citep[see e.g.][]{Rodgers.1984, Rubin.1986} in this paper which typically aims at a balanced usage of donors. We feel, however, that taking this approach to the extreme 'forces' the marginal distribution of $Z$ from the donor study upon the recipient study, irrespective of different sample properties, indicated by deviating distributions in $X$. Under these circumstances it would be plausible that the fused distribution of $\tilde{Z}$ in the recipient study should be different to the corresponding distribution in the donor study.

Besides, the primary objective of any data fusion is the preservation of the joint distribution of $Y$ and $Z$ \citep{Kiesl.2005} which might clash with the objective of preserving the marginal distribution of $Z$ at all costs. In our simulation studies PMM outperformed the covariate-based RHD method with respect to preserving $f(Y,Z)$.
While Predictive Mean Matching can only be applied to metric-scale $Z$ variables, we believe we have demonstrated that the method is a very useful addition to the toolbox of data fusion methods and, thus, should be taken into consideration for general application.

\section{Appendix}

\setlength{\parskip}{13pt}

\subsection*{Recoding Details of Certain $\boldsymbol{X}$ Variables}$\phantom{0}$

\noindent\underline{\textbf{Activity Status ($\boldsymbol{X_1}$)}}

\noindent Variable PL031 'Self-defined current economic status' \citep[p. 285]{Eurostat.2016} with the following levels:

\noindent {\phantom{0}1: Employee working full-time} \\ {\phantom{0}2: Employee working part-time} \\ {\phantom{0}3: Self-employed working full-time (including family worker)} \\ {\phantom{0}4: Self-employed working part-time (including family worker)} \\ {\phantom{0}5: Unemployed} \\ {\phantom{0}6: Pupil, student, further training, unpaid work experience} \\ {\phantom{0}7: In retirement or in early retirement or has given up business} \\ {\phantom{0}8: Permanently disabled or/and unfit to work} \\ {\phantom{0}9: In compulsory military community or service} \\ {10: Fulfilling domestic tasks and care responsibilities} \\ {11: Other inactive person}\newpage

\noindent
Grouping to \textit{activity status}:\\
$\phantom{0}$\hspace{6.5ex} \underline{Original category\phantom{:}} $\ \,\Longrightarrow$
\hspace{3ex}\underline{\textit{Activity status}}\smallskip\\
\renewcommand*{\arraystretch}{1}
$\phantom{0000000000000000001}\left.\begin{matrix}[r]
\ 1\\2\\3\\4
\end{matrix}\ \right\}\ \Longrightarrow \phantom{N}\;1$: Working\\$
\phantom{0000000000000000001}\left.\begin{matrix}\ 5
\end{matrix}\ \right.\ \ \  \Longrightarrow \phantom{N}\;2$: Unemployed
\\$
\phantom{0000000000000000001}\left.\begin{matrix}\ 7
\end{matrix}\ \right.\ \ \  \Longrightarrow \phantom{N}\;3$: In retirement or early retirement
\\$
\phantom{000000000000000000}\left.\begin{matrix}\ \,10
\end{matrix}\ \right.\ \ \,  \Longrightarrow \phantom{N}\;4$: Fulfilling domestic tasks
\\$
\phantom{0000000000000000001}\left.\begin{matrix}\ 8
\end{matrix}\ \right.\ \ \  \Longrightarrow \phantom{N}\;5$: Permanently disabled
\\
$\phantom{100000000000000000}\hspace{-0.4ex}\left.\begin{matrix}[r]
6\\9\\11\\\text{NA}
\end{matrix}\ \right\}\ \Longrightarrow \phantom{N}\;9$: Not specified (NA)
\renewcommand*{\arraystretch}{1.5}

\noindent The categories coded as 9 (NA) within the \textit{activity status} had a too small number of cases which causes problems in some MC samples.

\newpage
\noindent\underline{\textbf{Population Density Level ($\boldsymbol{X_3}$)}}

\noindent Random variable generation for Germany due to completely missing information in the German EU-SILC Scientific Usefile. We randomly generated the variable with a sampling probability that is analogue to the published distribution which can retrieved by the online Data Explorer \citep{Eurostat.2020}.

\noindent Variable DB100 'Degree of urbanisation' \citep[p. 112]{Eurostat.2016} with the following levels and its sampling probability:\\

\noindent
1: Densely-populated area (35.8\%)\\
2: Intermediate area (41.8\%)\\
3: Thinly-populated area (22.4\%)

\newpage
\noindent\underline{\textbf{Main Source of Income ($\boldsymbol{X_6}$)}}

\noindent The generated variable is based on six main sources of income that reflect for each individual the maximum total value of the following variables: 

\noindent
    PY010G: 'Employee cash or near cash income' \citep[p. 313]{Eurostat.2016}\\
    PY020G: 'Non-cash employee income' \citep[p. 315]{Eurostat.2016}\\
    PY050G: 'Cash benefits or losses from self-employment' \citep[p. 322]{Eurostat.2016}\\
    PY080G: 'Pension from individual private plans' \citep[p. 327]{Eurostat.2016}\\
    PY090G: 'Unemployment benefits' \citep[p. 328]{Eurostat.2016}\\
    PY100G: 'Old-age benefits' \citep[p. 328]{Eurostat.2016}\\
    PY110G: 'Survivor’ benefits' \citep[p. 328]{Eurostat.2016}\\
    PY120G: 'Sickness benefits' \citep[p. 328]{Eurostat.2016}\\
    PY130G: 'Disability benefits' \citep[p. 328]{Eurostat.2016}\\
    PY140G: 'Education-related allowances' \citep[p. 328]{Eurostat.2016} 
    
\noindent The variables are all metric scaled and the highest total value in each case leads to the following auxiliary categories:\\ \newpage\noindent

Maximum for\smallskip\\ \renewcommand*{\arraystretch}{1}
$\begin{matrix}[lcl] \text{-- PY010G or PY020G} & \Longrightarrow & \text{Wages or salary} \\ \text{-- PY050G} & \Longrightarrow & \text{Income from self-employment}\\ \text{-- PY080G} & \Longrightarrow & \text{Property income}\\ \text{-- PY100G} & \Longrightarrow & \text{Pensions}\\ \text{-- PY090G} & \Longrightarrow & \text{Unemployment benefits}\\ \text{-- PY110G, PY120G, PY130G or PY140G} & \Longrightarrow & \text{Other benefits}\\
\end{matrix}$\\

\noindent As long as all total values for the variables PY010G to PY140G amounts to zero, we code this as a missing value (NA). In this case, we included the missing values (NAs) and code it as 9.\\
Finally, the auxiliary categories are grouped into two categories that build the final variable \textit{main source of income} ($X_6$):

$\phantom{00008481\;}$\underline{Auxiliary category} $\quad\ \Longrightarrow\quad$ \underline{\textit{Main source of income}}

\renewcommand*{\arraystretch}{1}\noindent
$\phantom{1}\left.\begin{matrix}[r]
\ \text{Wages or salary}\\\text{Income from self-employment}\\\text{Property income}
\end{matrix}\ \right\} \Longrightarrow \phantom{N}\;1$\\

\noindent $\phantom{000011}\left.\begin{matrix}[r]
\ \text{Pensions}\\\text{Unemployment benefits}\\\text{Other benefits}
\end{matrix}\ \right\} \Longrightarrow \phantom{N}\;2$

\noindent$\phantom{000000000000000000000\,01}\left.\begin{matrix}[r]\,\text{NA}
\end{matrix}\ \right.\ \ \Longrightarrow \phantom{N}\;9$
\renewcommand*{\arraystretch}{1.5}

\newpage
\bibliographystyle{agsm}

\renewcommand{\baselinestretch}{1}

\bibliography{Literaturverzeichnis}

\end{document}